\documentstyle[pre,preprint,aps]{revtex}
\draft
\tighten

\begin{document}

\author{A.Borghardt and D.Karpenko \\
Physically Technical Institute of National 
Ukrainian Academy of Sciences, \\
Donetsk 83114, Ukraine}
\title{Thermodynamics of Bose-Einstein condensation \\ of relativistic gas}
\maketitle

\begin{abstract}
We show that the free relativistic wave equation which describes the
particle without or with rest mass has more than one part of energy
spectrum. One part of energy spectrum is beginning with rest energy and it
is not limited by above. This part of spectrum is called by us as normal.
Another part of energy spectrum is beginning with the zero energy . This
part of spectrum is called by us as anomalous since the zero energy
corresponds to the infinite group velocity.The presence of the zero in the
energy spectrum permits to consider the Bose-Einstein condensation. We show
that the heat capasity has the finite discontinuity at the condensation
temperature. The last means that we have the phase transition at the
condensation point.
\end{abstract}

\section{I. Introduction}

It is well known that the construction of the thermodynamics requires the
knowledge of particle energy spectrum because the principal thermodynamic
characteristic ,namely, equilibrium temperature distribution function
depends on particle energy (frequency). We denote the last by $f_r$ ,
\[
f_r=\left( \exp (\hbar \omega -\mu /k_BT)+r\right) ^{-1},r=0,\pm 1.
\]
where the relativistic energy $\hbar \omega $ is equal to

\begin{equation}
\hbar \omega =\hbar c\sqrt{k^2+k_0^2},
\end{equation}

Here $k^2$ is a square of 3-D momentum and $k_0$ is the inverse Compton wave
length. We have superhigh energy at $k>>k_0$, and in this case energy is
equal to $\hbar \omega \approx \hbar ck$ (ultrarelativistic case). In the
opposite case $k<<k_0$ the energy spectrum of Eq(1) transforms into non
relativistic energy spectrum. Energy spectrum of Eq(1) corresponds to
eigenfunctions (plane waves) of Klein-Gordon -Fock (KGF) equation
\begin{equation}
\Psi ^{(+)}=\exp \left( i(-\omega t+k_zz)\right) \exp i(k_1x_1+k_2x_2).
\end{equation}

Later we will explain the representation of the solution of Eq.(2) as a
product of two exponential functions . The plane waves of Eq.(2) are the
basis for Green functions (GF) of KFG equation in the time-like Minkowski
space $M^{(+)}$. It is known from long ago (see St\"uckelberg, Feynman,
Fierz [1-3]) that the construction of the causal relativistic propagator
(CRP) requires the knowledge of GF in space-like Minkowski space $M^{(-)}$
as well. Considering relativistic quantum mechanics in so-called
characteristic representation (CR) we can easily prove the last assertion
[4]. Let us discuss the main principles of relativistic dynamics of CR in a
brief form. In CR we have to single out one of the spatial coordinates in a
wave operator (for the sake of definiteness we assume it to be $z$ and let
us write the relativistic wave equation in the form of 1-D telegraphic
equation with the operator parameter $\widehat{a}^2$ ,
\begin{equation}
\left( \frac{\partial ^2}{\partial z^2}-\frac 1{c^2}
\frac{\partial ^2}{\partial t^2}+\widehat{a}^2\right) 
\Psi (x_1,x_2,z,t)=0.
\end{equation}
where 
\[
\widehat{a}^2=\frac{\partial ^2}{\partial x_1^2}+
\frac{\partial ^2}{\partial x_2^2}-k_0^2=
\Delta _{\bot }-k_0^2 
\]
is elliptic operator.

Dealing with CR it is convenient to classify states by the eigenfunctions of
the elliptic operator $\Delta _{\bot }$. This operator has two kinds of
solutions. The first solution is a plane wave. The last gives the
contribution to the GF in the time-like space $M^{(+)}$. The second solution
is the Macdonald function $K_0$ ,
\[
\Delta _{\bot }K_0(k_{\bot }x_{\bot })=k_{\bot }^2K_0(k_{\bot },x_{\bot }),
\]
where $x_{\bot }=\sqrt{x_1^2+x_2^2}$ and $k_{\bot }=\sqrt{k_1^2+k_2^2}$.

Therefore together with the plane wave there is one more solution
\begin{equation}
\Psi _{}^{(-)}=\exp 
\left( i(-\omega t+k_zz)\right) K_0(k_{\bot }x_{\bot }), 
\end{equation}
where $\omega $ is equal now to
\begin{equation}
\omega =c\sqrt{k_z^2-k_{\bot }^2+k_0^2}.
\end{equation}

We have shown in [5] that the both solutions of Eqs.(2) and (4) are the
basis for two GF. One is defined inside the light cone and another is out of
the light cone. It is especially simply to obtain two GF for photons when
rest mass is equal to zero. Taking into account that the 1-D telegraphic
equation (3) has the fundamental solution in the terms of Bessel function
(BF) of the first kind of order zero we obtain [4]
\[
G_{ph}^{(+)}=J_0\left( \sqrt{(c^2t^2-z^2)(-\triangle _{\bot })}\right)
\left| 0\right\rangle =\frac 1{2\pi }
\frac{\delta (\left| t\right| -r/c)}{cr}=\frac{\delta 
\left( r^2-c^2t^2\right) }{2\pi },
\]
where $\delta (z)$ is Dirac delta function, and here it's taken into account
that $\delta \left( \left| t\right| +r/c\right) =0$
and also
\[
G_{ph}^{(-)}=J_0\left( \sqrt{(z^2-c^2t^2)(\triangle _{\bot })}\right)
\left| \widetilde{0}\right\rangle =\frac 1{2\pi }\frac 1{(r^2-c^2t^2)}.
\]
Here
\[
\left| 0\right\rangle =\frac 1{4\pi ^2}\int 
\exp i(k_1x_1+k_2x_2)=\delta (x_1)\delta (x_2)
\]
and
\begin{equation}
\left| \widetilde{0}\right\rangle =\frac 1{2\pi }\int\limits_0^\infty
K_0(k_{\bot }x_{\bot })k_{\bot }dk_{\bot }=\frac 1{2\pi }
\frac 1{(x_1^2+x_2^2)} 
\end{equation}
are the initial values of GF which are given along 
the characteristics $\left| t\right| =\left| z/c\right| $. 
The sum of these GF is CRP $D_c$ [2]
\[
D_c=\frac 12G_{ph}^{(+)_{}}+\frac i\pi G_{ph}^{(-)}.
\]

The decomposition of the BF into the plane waves is presented in [5]. 
If rest mass is not equal to zero then the operator argument in the BF 
is changed: $-\triangle _{\bot }\rightarrow k_0^2-\triangle _{\bot }$. 
Inside the light cone we have the function $G^{(+)}=\Delta _s$, 
where $\Delta _s$ is the even
solution of KFG equation [6] and outside the light cone we have 
\[
G_{}^{(-)}=J_0\left( \sqrt{(z^2-c^2t^2)(\triangle _{\bot }^{}-k_0^2)}\right)
\left| \widetilde{0}\right\rangle =\frac{k_0K_1(k_0\sqrt{r^2-c^2t^2})}{
\sqrt {r^2-c^2t^2}} 
\]
where $\left| \widetilde{0}\right\rangle $ is the initial value of
propagator $G^{(-)}$ along characteristics $\left| t\right| =\left|
z/c\right| $ (compare with (6))
\begin{equation}
\left| \widetilde{0}\right\rangle =\frac 1{2\pi }\int\limits_{k_0}^\infty
K_0(k_{\bot }x_{\bot })k_{\bot }dk_{\bot }=\frac{k_0}{2\pi }
\frac {K_1(k_0x_{\bot })}{x_{\bot }}
\end{equation}

Let us write the group velocities along the determinate direction for energy
spectra of Eqs. (1) and (5).
For Eq.(1)
\[
v_z^{(+)}=\frac{\partial \omega }{\partial k_z}=
\frac{ck_z}{\sqrt {k_z^2+q_{\bot }^2+k_0^2}}<\left| c\right| ,
\]
for Eq.(5)
\[
v_z^{(-)}=\frac{\partial \omega }{\partial k_z}=
\frac{ck_z}{\sqrt {k_z^2-k_{\bot }^2+k_0^2}}\geq \left| c\right| .
\]
What differs the spectrum of Eq.(5) from the spectrum of Eq.(1)? The main
difference lies in the presence of minimal zeroes energy for which the
velocity $v_z^{(-)}$is maximal and goes to infinity. The zero energy
(frequency) is realized if the transverse momentum $k_{\bot }\neq 0$ and is
equal to
\[
k_{\bot }=\sqrt{k_z^2+k_0^2}\geq k_0$, at $\omega =0.
\]
Due to such extraordinary property that the infinitely- moving particle has
zero energy, the spectrum of Eq.(5) is called by us as anomalous,although
the definition of this spectrum as anomalous is not good. The state
describing this spectrum is needed as well as normal. As it has been
mentioned above, the CRP $D_c$consists of these two states (see Eqs.(2)and
(4)).

Energy spectrum of Eqs.(1) and (5) (dispersion curves) are mapped on the
plane $(\omega /c,k_z)$ by half-hyperbola which belong to different
quadrants of the plane $(\omega /c,k_z)$ divided by asymptotes $\left|
\omega /c\right| =\left| k_z\right| $ .

In this paper we examine the special features of thermodynamic
characteristics concerning statistical integral, internal and free energies,
enthropy etc. referring to relativistic gas in the anomalous state.

\section{Thermodynamic characteristics of the normal states of relativistic
gas}

For comparison we present in a brief form the known results of relativistic
thermodynamics (RT) based on the spectrum of Eq.(1) [$^7$] .First we
consider the Maxwell -Boltzmann(M-B) statistics , that is $r=0$ .We denote
the particle statistic integral in M-B statistics by $Z$
\begin{equation}
Z=\int d^3k\exp \left( -\frac{\hbar \omega }{k_BT}\right) =4
\pi k_0^3\int\limits_1^\infty 
\varepsilon \sqrt{\varepsilon ^2-1}\exp \left( 
\frac{-\Theta \varepsilon }T\right) d\varepsilon ,
\end{equation}
where $\Theta =mc^2/k_B$ is typical parameter of RT having the temperature
dimensionality and $k_B$ is Boltzmann constant. If $m$ is electron mass then 
$\Theta \approx 10^{10^{\circ }}K.$

Let us denote the factor before the exponential function in integral of Eq.
(8) by $g(\varepsilon )$
\[
g(\varepsilon )=4\pi k_0^3\varepsilon \sqrt{\varepsilon ^2-1},
\]
or in dimensional units ($\varepsilon =\hbar \omega /mc^2$)

\begin{equation}
g(\omega /c)=4\pi k_0^3\frac{\hbar \omega }{mc^2}
\sqrt{\left( \frac{\hbar
\omega }{mc^2}\right) ^2-1} . 
\end{equation}

Eq.(9) gives the state density in the energy scale which is determined up to
the factor $\frac \Omega {8\pi ^3},$where $\Omega $ is the volume of
relativistic gas. This density is equal to zero if $\hbar \omega \leq mc^2$
or in temperature units $T\leq \Theta $.Due to such temperature restriction
it is said that RT has academic interest and it can be applicable in
astrophysical phenomena where there are superhigh temperatures and
densities. The integral of Eq.(8) is easily calculated by 
substitution $\varepsilon =\cosh \tau $ and with the help of 
the table integral $\int\limits_0^\infty dz\exp (-q\cosh z)=K_0(q)$. 
It is respectively equal to
\begin{equation}
Z=4\pi k_0^3\frac{K_2(\frac \Theta T)}{(\frac \Theta T)}
\end{equation}
where $K_2$ is the Macdonald function of the second kind.

Let us calculate the asymptotics of Eq.(10) at low temperature when $\frac
\Theta T>>1$
\[
Z\approx (2\pi )^{3/2}k_0^3(\frac \Theta T)^{\frac 32}\exp 
\left( -\frac \Theta T\right) 
\]
Statistic integral of Eq.(10) does not depend on mass at high temperature
when $(\frac \Theta T)<<1$
\[
Z\approx 8\pi k_0^3(\frac T\Theta )^3=8\pi 
\left( \frac{k_BT}{\hbar c}\right) ^3.
\]
Let us compute the internal energy density
\[
U=\left\langle \hbar \omega \right\rangle =\int d^3k\hbar \omega \exp
\left( \frac{-\hbar \omega }{k_BT}\right) =k_BT^2\frac{
\partial Z}{\partial T } .
\]
It is convenient to find the internal energy in per one particle so-called
middle energy
\[
\overline{U}=\frac{\left\langle \hbar \omega \right\rangle }Z=k_BT^2\frac
\partial {\partial T}\ln Z=\mu c^2(\frac T\Theta +\frac{K_1+K_3}{2K_2}).
\]
At low temperature the middle energy and heat capacity (HC) are equal to
\[
\overline{U\approx }\mu c^2+3/2k_BT,$ $c_v=\frac 32k_B.
\]
The middle energy and HC at high temperature $(\frac \Theta T)<<1$ are equal
to
\[
U\approx 3k_BT,$ $c_v=3k_B.
\]
If the relativistic gas obeys to Bose-Einstein (B-S) statistics 
that is $r=-1 $ then all thermodynamic characteristics can be 
found from the distribution
\[
N=4\pi k_0^3\sum\limits_{n=1}^\infty 
\frac{\exp (\mu n/k_BT)K_2(n\Theta /T) }{(n\Theta /T)}.
\]

\section{Thermodynamics of anomalous states}

Let us find the state density in an energy scale for the energy spectrum of
Eq.(5). It is convenient to introduce dimensionless variables, namely,
dimensionless energy (frequency) and dimensionless momentum variables
\[
\epsilon =\omega /k_0c,$ $q_z=k_z/k_0,$ $q_{\bot }=k_{\bot }/k_0.
\]
We write dispersion equation (5) in dimensionless variables
\[
\varepsilon =\sqrt{q_z^2-q_{\bot }^2+1}.
\]
In fact, it is necessary to make the transition from the momentum
representation to energy representation with the help of transformation
formulae which depend on the connection between full 
energy ($\hbar \omega $ ) and rest energy ($mc^2$). 
It means that in dimensionless variables the value 
 $\varepsilon $ can be more than one or less than one. 
Let $\varepsilon >1$ then transformation formulae have the form
\[
\begin{array}{c}
q_z=\sqrt{\varepsilon ^2-1}\cosh v,\text{ }q_1=\sqrt{\varepsilon ^2-1}\cos
u\sinh v,q_2=\sin u\sinh v.
\end{array}
\]
The infinitely -small volume element $d^3k$ ($d^3q$ in dimensionless
variables ) in the new variables has the form

\begin{equation}
d^3q=\left| I\right| d\varepsilon dudv ,
\end{equation}

where $I$ is Jacobian of transformation . 
It is equal to $I=\epsilon \sqrt {\varepsilon ^2-1}\sinh v.$

To obtain the state density in energy scale it is needed to integrate
Eq.(11) over two angles $u$ and $v$ .The integration over the angle $u$
gives the factor $2\pi $. To integrate Eq.(11) over hyperbolic angle $v$ it
is necessary to know its limits of integration. They are found from formula
for the domain of the definition of transverse 
momentum $k_{\bot }$ ( $q_{\bot }$in dimensionless variables )

\begin{equation}
q_{\bot }=\sqrt{q_1^2+q_2^2}=\sqrt{\varepsilon ^2-1}\sinh v .
\end{equation}

The low limit of the hyperbolic angle $v$ corresponds to the 
minimal value $q_{\bot }$. Since $q_{\bot }^{\min }=1$(or in 
dimension variables $k_{\bot }^{\min }=k_0$(see Eq.(7)) then
\[
\sinh v_{\min }=\frac 1{\sqrt{\varepsilon ^2-1}}, 
\cosh v_{\min }=\sqrt{ 1+\sinh ^2v_{\min }^{}}=\frac 
\varepsilon {\sqrt{\varepsilon ^2-1}} .
\]
If the upper limit of the value $q_{\bot }$ is equal to infinity then the
upper limit of hyperbolic angle $v$ is equal to infinity too. In this case
Eq.(11) becomes infinite after integration over hyperbolic angle $v$
.Therefore we cut the momentum $k_{\bot }$ by the 
value $k_{\bot }^{\max }.$
It does not influence on the state with zero energy because
\[
\omega =0$ at $k_{\bot }^{\max }=
\sqrt{(k_z^{\max })^2+k_0^2}>k_0$ or $q_{\bot }^{\max }>1.
\]
We can find the upper integral limit from Eq.(12)
\[
\sinh v_{\max }=\frac{q_{\bot }^{\max }}{\sqrt{\varepsilon ^2-1}}, 
\cosh v_{\max }=\frac{\sqrt{(q_{\bot }^{\max })^2+\varepsilon ^2-1}}{
\sqrt {\varepsilon ^2-1}} .
\]
Integrating Eq.(11) over the angles $u$ and $v$ we obtain the state density
in the energy scale

\begin{equation}
g(\varepsilon )d\varepsilon =2\pi k_0^3\left( 
\varepsilon \sqrt{\varepsilon
^2+(q_{\bot }^{\max })^2-1}-\varepsilon ^2\right) d\varepsilon 
\end{equation}

Further we consider the case $\varepsilon <1$. It means that in the
dimensional variables energy is less than rest 
energy $mc^2$ $ ($or $\varepsilon <1)$. 
In this case the transformation formulae have the form
\[
\begin{array}{c}
q_z= \sqrt{1-\varepsilon ^2}\sinh v\text{ , }q_1=
\sqrt{1-\varepsilon ^2}\cos u\cosh v\text{ ,} \\ 
\text{ }q_2=\sqrt{1-\varepsilon ^2}\sin u\cosh v\text{ ,}
\end{array}
\]
and the volume element in dimensionless variables has the form

\begin{equation}
d^3q=\left| I\right| d\varepsilon dudv=\varepsilon 
\sqrt{1-\varepsilon ^2} d\varepsilon dud(\sinh v)
\end{equation}

where $I=\varepsilon \sqrt{1-\varepsilon ^2}\cosh v$ is Jacobian of
transformation.

We can find the limits of the angle $v$ from the expression
\[
q_{\bot }=\sqrt{q_1^2+q_2^2}=\sqrt{1-\varepsilon ^2}\cosh v.
\]
Repeating preceding reasonings we find
\[
\cosh v_{\min }=\frac 1{\sqrt{1-\varepsilon ^2}}\,, 
\sinh v_{\min }=\frac \varepsilon {\sqrt{1-\varepsilon ^2}} .
\]
and
\[
\cosh v_{\max }=\frac{q_{\bot }^{\max }}{\sqrt{1-\varepsilon ^2}}\,,\, 
\sinh v_{\max }=\frac{\sqrt{(q_{\bot }^{\max })^2+
\varepsilon ^2-1}}{\sqrt {1-\varepsilon ^2}} .
\]
Performing in Eq.(14) the integration over angle variables we find the state
density coinciding with Eq.(13).

Let us rewrite Eq.(13) in dimensional variables

\begin{equation}
\frac{g(\omega /c)}{k_0}=2\pi 
\left( \frac \omega c\sqrt{\frac{
\omega ^2}{c^2}+(k_{\bot }^{\max })^2-k_0^2}-
\frac{\omega ^2}{c^2}\right) 
\end{equation}

If the value $k_0$ in Eq.(15) goes to zero then the form of this equation is
not changed. From Eq.(15) it follows
\[
\lim g(\omega /c)=\pi k_0
\left( (k_{\bot }^{\max })^2-k_0^2\right)\,\,  at
\,\,\omega \rightarrow \infty  .
\]
Introducing the notion of effective 
mass $m^{*}=m\sqrt{(q_{\bot }^{\max })^2-1}$, 
we rewrite the state density of Eq.(13) in a more compact form

\begin{equation}
g(\omega /c)=2\pi k_0^3\left( \varepsilon 
\sqrt{\varepsilon ^2+\left( 
\frac {m^{*}}m\right) ^2}-\varepsilon ^2\right) 
\end{equation}

Now we can consider the statistic integral of anomalous state in M-B
statistics in energy representation 
\begin{equation}
Z=2\pi k_0^3\int\limits_0^\infty \left( \varepsilon \sqrt{\varepsilon
^2+\left( \frac{m^{*}}m\right) ^2}-
\varepsilon ^2\right) \exp \left( -\frac {\Theta 
\varepsilon }T\right) d\varepsilon  
\end{equation}
One integral in Eq.(17) is calculated in an 
elementary manner 
\[
\int\limits_0^\infty \varepsilon ^2\exp 
\left( -\frac{\Theta \varepsilon } T\right) d
\varepsilon =2\left( \frac T\Theta \right) ^3
\]
The second integral
of Eq.(17) is calculated with the help of 
substitution $\varepsilon =\frac{m^{*}}m\sinh \theta $ and 
of table integral $\int\limits_0^\infty d\theta $  
 $\exp (-q\sinh \theta )=\frac \pi 2A_0(q)$, 
where $A_0={\bf H}_0-N_0$ and ${\bf H}_0$ is Struve function 
and $N_0$ is the Bessel function of the second
kind of the order zero. 
From what has been said above it follows that 
\begin{equation}
Z=\pi ^2\left( \frac{m^{*}c}\hbar \right) B\left( 
\frac{\Theta ^{*}}T\right) 
\end{equation}
where $\Theta ^{*}=\frac{m^{*}c^2}{k_B}= $  
 $\frac{mc^2}{k_B}\sqrt{(q_{\bot }^{\max })^2-1}$ 
is effective temperature and 
\begin{equation}
B(q)=2A_1/q^2-A_0/q-4/\pi q^3\text{, }A_i={\bf H}_i-N_i,
\text{ }i=1,2  
\end{equation}
The spectrum of Eq.(5) unlike the spectrum of Eq.(1) 
permits to consider B-E
condensation of relativistic gas. In this case the cutting 
parameter $k_{\bot }^{\max }$ (or effective mass $m^{*}$) 
can be expressed in the terms
of condensation temperature $T_c$ and gas particle number $N_a^{}$ per unit
volume. The condensation temperature $T_c$ is defined from the condition of
vanishing of chemical potential $\mu $ in 
the gas particle density $N_a^{}.$
From the point of view of B-E statistics the $N_a$ is equal to 
\begin{equation}
N_a=\int\limits_0^\infty 
\frac{g(\varepsilon )d\varepsilon }{\exp \left( 
\frac{\Theta \varepsilon }{T_c}\right) -1}=
\pi ^2(k_0^{*})^3\sum\limits_{n=1}^\infty B
\left( \frac{n\Theta ^{*}}{T_c} \right)  
\end{equation}

where $k_0^{*}=\frac{m^{*}c}\hbar $ and the value $B(q)$ is defined from
Eq.(19). At low temperature the 
condition $\frac{n\Theta ^{*}}{T_c}>>1$ is
always performed and we can use the asymptotics of the 
function $B(q)\approx 2/\pi q^2$ at $q>>1$ ,

\begin{equation}
N_a\approx 2\pi \varsigma (2)(k_0^{*})^3\left( 
\frac{T_c}{\Theta ^{*}}\right) ^2=2\pi 
\varsigma (2)k_0^{*}\left( \frac{k_BT_c}{\hbar c}\right) ^2
(21)
\end{equation}
where $\varsigma (s)$ is zeta-function of Riemann .

From Eq.(21) it follows that the effective mass $m^{*}$ satisfies the
condition 
\[
m^{*}T_c^2=const. 
\]
The gas particle density $N_0^{}$with the zero energy (condensation
particles) and the density of overcondesation 
particles $N_1$of relativistic
gas are equal to 
\[
N_0=N_a\left( 1-\left( \frac T{T_c}\right) ^2\right) ,
\text{ }N_1=N_a\left(
\frac T{T_c}\right) ^2. 
\]
The internal energy density $U$ of relativistic gas in temperature 
interval $0\leq T\leq T_c$ is defined by integral 
\begin{equation}
U=mc^2\int\limits_0^\infty 
\frac{\varepsilon g(\varepsilon )d\varepsilon }{\exp 
\left( \frac{\Theta \varepsilon }T\right) -1}=
\pi ^2k_0^{*3}m^{*}c^2\sum\limits_{n=1}^\infty D\left( 
\frac{n\Theta ^{*}} T\right) 
\end{equation}
where $D(q)=2/\pi q-12/\pi q^4-A_1/q-3A_0/q^2+6A_1/q^3$.

Taking into account that at low temperature ($q>>1$) the 
asymptotics of $ D(q)-$function is$\sim 4/\pi q^3$ we have from Eq.(22) 
\begin{equation}
U\approx 4\pi \varsigma (3)k_0^{*3}m^{*}c^2\left( 
\frac T{\Theta
^{*}}\right) ^3\cong 1,5N_ak_BT\left( 
\frac T{T_c}\right) ^2 
\end{equation}
In Eq. (23) the approximate equality has been 
used $2\varsigma (3)/\varsigma (2)\cong 1,5.$

The HC $C_v$ in the temperature interval $0\leq T\leq T_c$ is equal
respectively to
\[
C_v=4,5N_ak_B\left( \frac T{T_c}\right) ^2.
\]

Let us investigate the relativistic thermodynamics in the temperature
interval $T-T_c<<T_c.$ We take into account that $N_0$ is equal 
to zero and $\left| \mu \right| /k_BT<<1.$ 
Then the particle density in relativistic gas
has the form 
\begin{equation}
N_a=\int\limits_0^\infty 
\frac{g(\varepsilon )d\varepsilon }{\exp \left( 
\frac{\Theta (\varepsilon +\mu /mc^2)}T\right) -1}
\simeq 2\pi (k_0^{*})_{}^3\varsigma (2)\left( 
\frac T\Theta \right) ^2\exp \left( \frac {-\mu }{k_BT}\right) 
\end{equation}

In Eq.(24) we have used the asymptotics of $B(q)$ and the approximate
summation with the help of Euler 
formula $\sum\limits_{n=1}^\infty \frac{\exp (-nq)}{n^2}$ 
 $\approx \varsigma (2)\exp (-q)$ which is valid at $q<<1$.

Substituting Eq.(21) in the left side of Eq.(24) 
we obtain the equation for the chemical potential $\mu $ ,
\[
\left( \frac T{T_c}\right) ^2=
\exp (\mu /k_BT)\rightarrow \mu =2k_BT\ln
\left( \frac T{T_c}\right) \text{ at }T\geq T_c. 
\]

All the derivatives of chemical potential $\mu $ are discontinuous at the
condensation temperature $T=T_c$, in contrary to non relativistic
thermodynamics where the first derivative of chemical potential $\mu $ is
continuous at the condensation temperature $T=T_c$ ,
\[
\mu ^{^{\prime }}(T_c-0)=0\,\,,\,
\mu ^{^{\prime }}(T_c+0)=2k_B \,,
\]
and etc.

Let us calculate the internal energy 
density $U$ in the temperature interval $T\geq T_c$ ,
\[
\begin{array}{c}
U=mc^2\int\limits_0^\infty \frac{\varepsilon g(\varepsilon )d
\varepsilon }{\exp \left( 
\frac{\Theta (\varepsilon +\mu /mc^2)}T\right) -1}= \\ 
4\pi k_0^{*3}m^{*}c^2\varsigma (3)
\left( \frac T{\Theta ^{*}}\right) ^2\exp
\left( -\frac \mu {k_BT}\right) \simeq 1,5N_ak_BT\text{,}
\end{array}
\]
and HC is equal to $C_v=1,5N_ak_B$. Thus, in the temperature 
interval $T\geq T_c$ the internal energy and HC 
of relativistic gas in the anomalous state
coincide with energy and HC of ideal non relativistic gas.

Let us write the expression for full internal energy using the 
Heaviside step-function $\theta (t)$ ,
\[
\begin{array}{c}
U(T)=1,5N_ak_BT\left\{ \left( 
\frac T{T_c}\right) ^2\theta (T_c-T)+\theta
(T-T_c)\right\} \text{ ,} \\ 
U(T_c)=1,5N_ak_BT_c\text{ .}
\end{array}
\]
At the condensation temperature $T=T_c$ we have the heat capacity jump 
\begin{equation}
\frac{C_v(T_c-0)-C_v(T_c+0)}{C_v(T_c+0)}=2  
\end{equation}
Eq.(25) does not contain the arbitrary parameters and therefore it admits
the experimental verification.

Enthropy variation can be found from the expression $dS=\frac{C_vdT}T$
whence , after integration we obtain
\[
S(T)=\frac 94k_BN_a\left\{ \left( 
\frac TT\right) ^2\theta (T_c-T)+\left(
\ln \left( \frac T{T_c}\right) ^{2/3}+1\right) 
\theta (T-T_c)\right\} 
\]
and $S(T_c)=\frac 94N_ak_B$.

We can find the expressions for free energy $F=U-TS,$ for 
an enthalpy $H=U+PV=TS+\mu N_a\theta (T-T_c),$ 
the equation of state of relativistic gas
etc. These main values have the fracture (or angle point) at the
condensation temperature $T_c.$ 
It means that the derivative of these values
with respect to temperature $T$ has the discontinuity of 
the first kind at $T=T_c$. Therefore 
these thermodynamic characteristics cannot be expanded into
Taylor series in the vicinity of the critical point $T=T_c.$

For comparison let us write the particle density $N_{n,r}$ of non
relativistic gas at the condensation temperature $T_c$ ,
\[
N_{n,r}=(2\pi )^{3/2}\varsigma (3/2)k_0^3
\left( \frac{T_c}\Theta \right)^{3/2} 
\]

The ratio of two densities (see Eq.(21)) 
\[
\frac{N_a}{N_{n,r}}=\frac{\varsigma (2)}{\sqrt{2\pi }
\varsigma (3/2)}\frac {m^{*}}m\left( \frac{T_c}\Theta 
\right) ^{1/2}\text{ } 
\]

shows that at low temperature $T_c$ the numbers $N_a<<N_{n,r}$ . 
This inequality strengthens if $m^{*}<<m.$

\section{Conclusion}

Our reasonings are at least in harmony with the papers of Bilaniuk,
Terletskiy , Feinberg (see ref.[8] and refs. therein ) in which it is proved
that the relativistic quantum mechanics is incomplete if we are restricted
by energy spectrum of Eq.(1). We know that the RCP $D_c$ consists of the
solutions of different nature (oscillatory and exponential damping). The
last corresponds to energy spectrum of Eq.(5) which we call anomalous. We
could call this spectrum as tachyon although tachyon term is referred to the
solutions of KGF equation with imaginary mass[9]. We believe that the energy
spectrum of Eq.(5) can be applied to the description of thermodynamics of
models with the observed phase transition of the second kind.

\,\,

\,\,

{\bf REFERENCES}

\,\,

1 E. St\"uckelberg, Helv.Phys.Acta, {\bf 14}, 588, (1941) 

2 R.Feynman, Phys.Rev., {\bf 76}, 769 (1949)

3 M.Fierz , Helv.Phys.Acta, {\bf 23}, 731 (1950) 

4 A. Borghardt and D. Karpenko , J. Nonlinear Math. Phys., 
{\bf 5}, 357 (1998)

5 A. Borghardt, D.Karpenko, JETP, {\bf 85}(4), 635, (1997)

6 N.Bogoliubov and D. Shirkov, {\it Introduction to the} 
{\it Theory of Quantized Fields} (Interscience, NY,1959)

7 W.Pauli, {\it Relativit\"atstheorie}, 
Ens. der Math. Wiss., {\bf 5}, 539 (1921)

8 O.Bilaniuk, V.Deshpande, E.Sudarshan, Amer.J.Phys., 
{\bf 30}, 718,(1962)

9 G.Feinberg, Phys.Rev.,{\bf 159}, 1089 (1967)


\end{document}